# Connectivity-Dependent Attenuation Factor in Nanographene-Based Molecular Wires

Sara Sangtarash

sara.sangtarash@warwick.ac.uk

Device Modelling Group, School of Engineering, University of Warwick, CV4 7AL Coventry, United Kingdom,

## Abstract

Designing molecular nanowires with high electrical conductance that facilitate efficient charge transport over long distances is highly desirable for future molecular-scale circuitry. However, most molecular wires act as tunnel barriers, and their electrical conductance decays exponentially with increasing length. Only recently have a few studies shown increasing conductance with length. In this study, we identify a new class of molecular wires that exhibit both an increase and a decrease in room-temperature conductance with length (a dual attenuation factor), depending on their connection points to the electrodes. We show that this dual attenuation factor is an inherent property of these graphene-like nanowires, and its demonstration depends on the constructive quantum interference pattern for different connectivities to the electrodes. This is significant because a given nanographene molecular wire can show both negative and positive attenuation factors. This enables the systematic design of connectivity-dependent high/low-conductance molecular wires for future molecular-scale circuitry.

## Introduction

Understanding electron transport in molecular junctions, identifying length dependence and achieving long-range charge transport across individual molecules are important for the advancement of molecular electronics.[1–5] While a wide variety of molecular nanowires have been studied for many years, most of them typically act as tunnel barriers and their conductance (G) decays exponentially with molecular length ($L$) as $G = Ae^{-\beta L}$ where $A$ is the prefactor and $\beta$ is the decay (attenuation) factor.[3,6] This exponential decay in conductance with wire length severely limits their applications for future molecular-scale circuitry. For example, conjugated molecular wires such as oligophenylene[7] show conductance values that decay with an increasing number of phenyl units, with $\beta$ = 0.41 Å$^{-1}$. For oligo(phenylene-ethynylenes) OPEs, measured room-temperature values of $\beta$ range from 0.2–0.34 Å$^{-1}$,[8] $\beta$ = 0.33 Å$^{-1}$ for oligo(aryleneethynylenes) OAEs.[9] Other molecular wires such as oligophenyleneimine[10], oligonaphthalenefluoreneimine[11] and alkanedithiol[3] showed attenuation factors of 0.3 Å$^{-1}$, 0.25 Å$^{-1}$ and 0.9 Å$^{-1}$ respectively. Recently there has been growing interest in the design of molecular nanowires with increasing conductance with length, called negative attenuation ($\beta$) molecular wires. As an example, negative $\beta$ has been demonstrated in cumulene and fused porphyrin molecular wires[12–19].

In the molecular structures studied today, a particular molecular wire exhibits either a positive or negative attenuation factor. In this paper, for the first time, we identify a new class of nanographene molecular wires where a given molecular wire shows both positive and negative attenuation factors (a dual attenuation factor, or DAF), depending on the connection point to the electrode. This connectivity-dependent DAF arises from room-temperature quantum interference in the multipath nanographene structures. In what follows, we will first demonstrate the DAF effect using a simple tight-binding method[20,21] and confirm it using material-specific first-principles calculations.

## Results and discussion

Figure 1A shows the molecular junction formed from a nanographene molecular core connected to electrodes via two different connections, e.g., (i,j) and (i,k). To simplify the notation, we use "blue connectivity" for the (i,j) connection to the electrodes and "red connectivity" for the (i,k) connection. The nanographene molecular wire (NMW) consists of a naphthalene unit (n in Fig 1A) connected to neighboring naphthalene units to form wires of different lengths. We first construct the tight-binding Hamiltonian of these structures with lengths from n=1 to n=20 naphthalene units, using $\varepsilon$=0 eV for the on-site energies and $\gamma$=-2.7 eV for the coupling integrals between neighboring sites[22]. We then calculate the transmission coefficient of electrons with energy $E$ passing from one electrode to the other through the molecule connected to the electrodes with the blue and red connectivities. The conductance can then be calculated from the Landauer formula $G = G_0 T(E_F)$, where $G_0$ is the conductance quantum and $E_F$ is the Fermi energy of the electrodes.

Our calculations show that, depending on the connection point to the electrodes, the conductance of the NMW increases (negative attenuation factor) or decreases (positive attenuation factor) with length. For junctions formed through the red connectivity (Figure 1), the conductance decreases with the length of the NMW around E=0 eV in Figure 2A. In contrast, the conductance surprisingly increases with the length of the NMW around E=0 eV for NMWs with the blue connectivity to the electrodes, as shown in Figure 2B, leading to a negative attenuation factor. The positive and negative attenuation factors are intrinsic properties of the NMW, which arise from the constructive quantum interference (CQI) pattern in this molecular core. This result is very significant because it shows that a given molecular wire can exhibit conventional behavior (a decrease in conductance with length) if connected through one (e.g., red) connectivity, but surprisingly different behavior (an increase in conductance with length) by connecting it to the electrode through another (e.g., blue) connectivity.

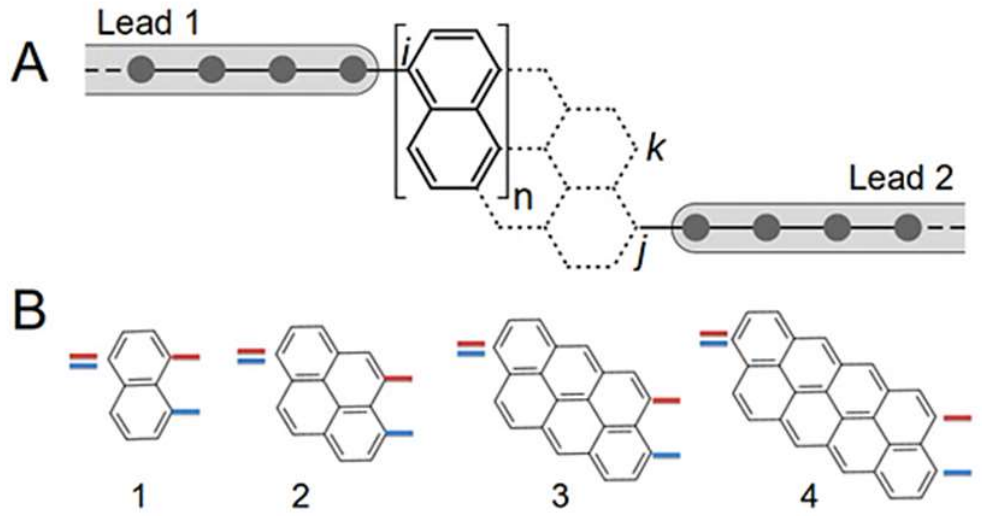

**Figure 1**: Molecular structure of naphtalene molecular wire (NMW). A) NMW between two 1D electrodes, B) Examples of the NMW with length 1 to 4. The red and blue lines represent the connection points to the electrodes.

Figure 2C shows the changes of transmission for molecular wire shown in Figure 1A for lengths up to 20 naphthalene units (n=20). The red (blue) curve represents changes of transmission coefficient for red (blue) connectivites to electrode. Clearly, the conductance increase rapidly initially and then saturates for blue connectivity whereas it decays for red connectivity.

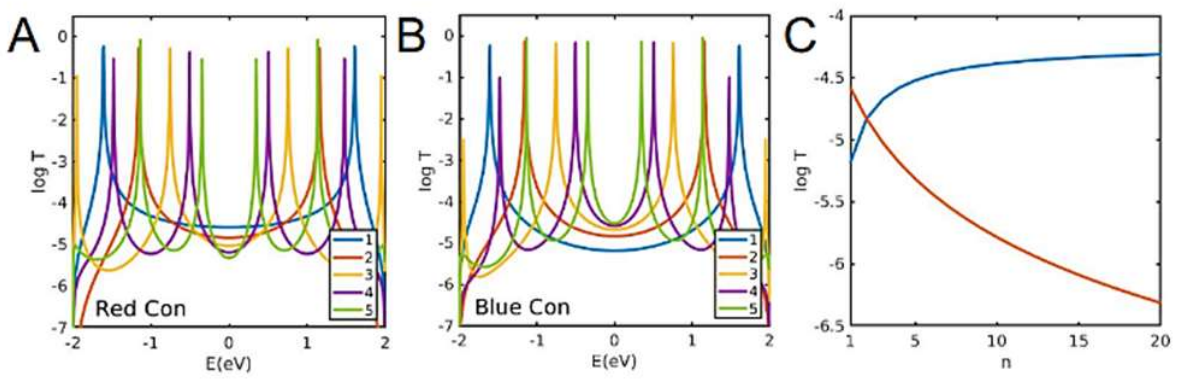

**Figure 2:** Tight-binding transmission coefficient *T*. A and B) Transmission coefficient of red and blue connectivities of the molecules 1- 5 respectively (structures shown in Figure 1). C) Transmission coefficient versus length of NMW at E=0eV demonstrating dual beta factor in NMW junctions.

To confirm the result obtained from the simple tight-binding model, which considers the contribution of pi orbitals only, we also performed material-specific calculations using density functional theory (DFT). We obtained the material-specific Hamiltonian of junctions formed by the NMW core 1-5 (see Figure 1) connected to two gold electrodes through thiol anchors and acetylene linkers. Figures 3A and 3C show examples of such junctions for NMW with a length of 5 naphthalene units. We then calculated the transmission coefficient of electrons with energy

*E* traversing from the right gold electrode to the left one using our quantum transport code Gollum[23]. Figures 3B and 3D show the corresponding transmission coefficient for two different connectivities and the same molecular core versus different lengths. For a connection similar to the blue connectivity in Figure 1, DFT transmission calculations show very similar behavior, where the conductance increases with length from 1 to 5 (Figure 3B). Crucially, this happens for all energies between the HOMO and LUMO gaps, confirming the robustness of the negative attenuation factor effect in these junctions. In contrast, when the same molecular core is connected through the red connectivity (similar to the red connectivity in Figure 1), the conductance decreases with length from 1 to 5 for a wide range of energies around the DFT Fermi energy (E=0). These material-specific calculations confirm the DAF effect in these molecular cores.

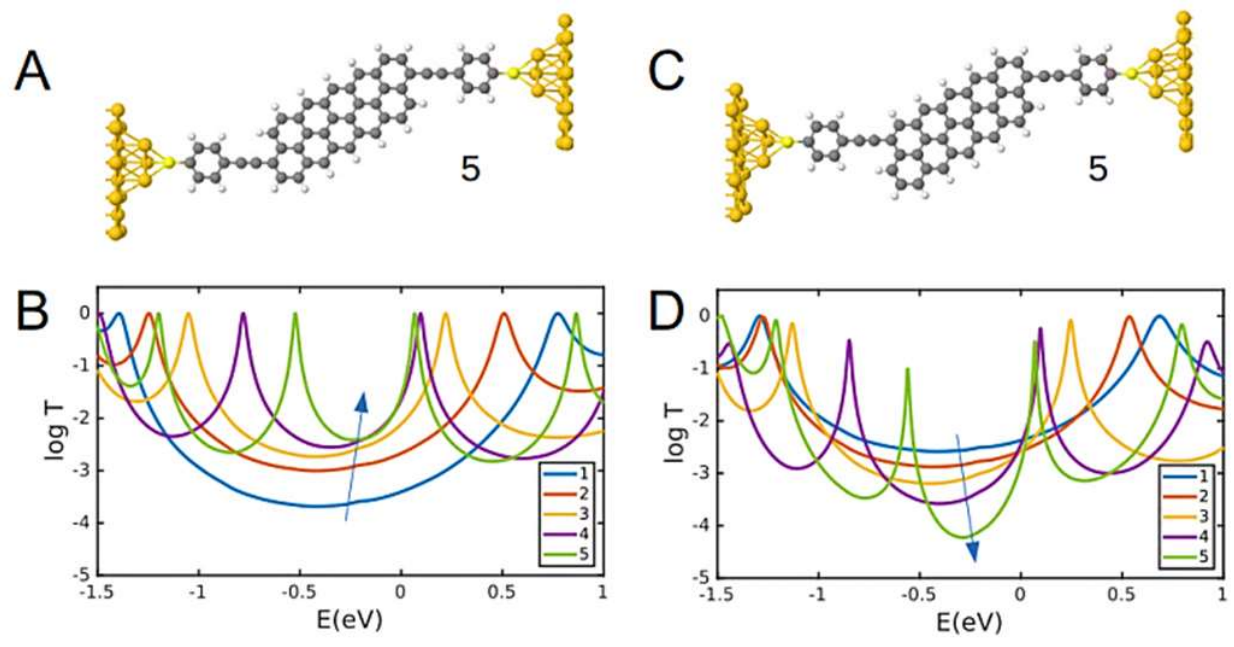

**Figure 3:** Transmission coefficient from material specific DFT-NEGF calculations. A and C) Example of junctions formed using NMW core with length 5 connected to gold electrodes through thiol anchors and acetylene linkers. B and D) Transmission coefficient for junctions with NMW cores and connectivites shown in A and C, respectively with lengths between 1-5.

To further investigate whether the DAF is unique to only the red and blue connectivities, we performed tight-binding calculations with other connectivities, as shown in Figure 4A. We found that similar behavior is observed for the connectivities shown in green and orange in Figure 4A. Figure 4B shows the transmission coefficient through molecules 1-4 connected to the electrodes with the green connectivity. The conductance decreases with length around E=0 eV for this connectivity. In contrast, the conductance increases with length when the molecule is connected to the electrodes through the orange connectivity (Figure 4C). Similar to the red and blue connectivities (Figure 2C) of NMW to the electrodes, the electron transmission at E = 0 eV increases rapidly with length initially and then saturates for the NMW with the orange connectivity to the electrodes, whereas it decays with length for the green connectivity.

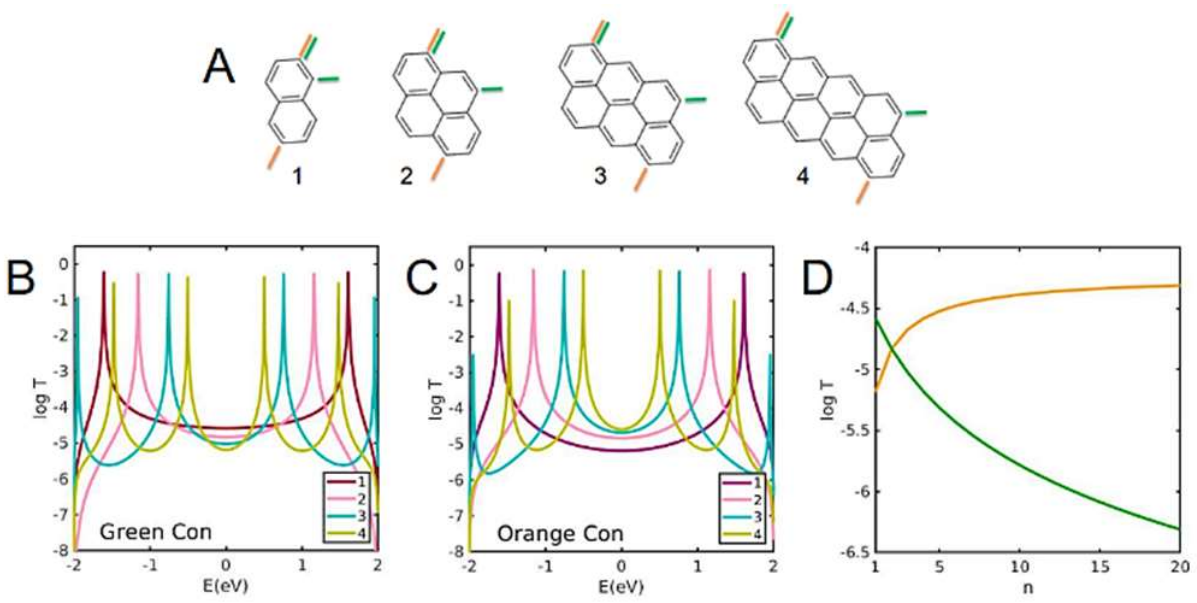

**Figure 4:** NMW with another set of connectivities to electrodes. A) Molecular structures and connection points (green and orange) to electrodes for NMWs with length 1 to 4. B and C) Transmission coefficient of the molecular wires with green and orange connectivities, respectively from length 1-4. D) Transmission coefficient versus length of NMW at E=0eV demonstrating dual beta factor in NMW junctions with green and orange connection points to electrodes.

## Conclusions

In summary, we demonstrated that a given molecular core (e.g., NMW) can exhibit two different attenuation factors, depending on the connection point to the electrodes and the constructive quantum interference pattern formed. We call this the DAF effect and demonstrate it using a simple tight-binding model as well as material-specific DFT-NEGF calculations. This enables the systematic design of connectivity-dependent high/low-conductance molecular wires for future molecular-scale circuitry.

## Theoretical methods

The Hamiltonian of the structures described in Figures 2 and 4 was constructed using a tight-binding approximation with a single orbital per atom, a site energy ε = 0 for carbon, and nearest-neighbor couplings γ = −2.7 for both C—C bonds.

**Geometry optimization:** The geometry of the structures shown in Figure 3 was relaxed to a force tolerance of 10 meV/Å using the SIESTA[24] implementation of density functional theory (DFT). A double-ζ polarized basis set (DZP) and the Local Density Approximation (LDA) functional with CA parameterization were used. A real-space grid was defined with an equivalent energy cut-off of 250 Ry.

**Transport Calculation:** A simple tight-binding Hamiltonian or mean-field Hamiltonian H obtained from the converged DFT calculation were combined with our implementation of the nonequilibrium Green's function method, Gollum,[23] to calculate the phase-coherent, elastic scattering properties of each system consist of left (source) and right (drain) leads and the scattering region (molecule). The transmission coefficient [6] $\mathrm{T(E)}$ for electrons of energy E (passing from the source to the drain) is calculated via the relation $\mathrm{T(E)} = \mathrm{Trace}\left(\Gamma_R(E) G^R(E) \Gamma_L(E) G^{R\dagger}(E)\right)$. In this expression, $\Gamma_{L,R}(E) = i\left(\sum_{L,R}(E) - {\sum_{L,R}}^{\dagger}(E)\right)$ describe the level broadening due to the coupling between left (L) and right (R) electrodes and the central scattering region, $\sum_{L,R}(E)$ are the retarded self-energies associated with this coupling and $G^R = (ES - H - \sum_L - \sum_R)^{-1}$ is the retarded Green's function, where H is the Hamiltonian and S is the overlap matrix.

**Electrical conductance:** Using the obtained transmission coefficient ($\mathrm{T(E)}$), the conductance could be calculated by the Landauer formula ($G = G_0 \int dE\, T(E)(-\partial f/\partial E)$) where $G_0 = 2e^2/h$ is the conductance quantum, $f(E) = (1 + \exp((E - E_F)/k_B T))^{-1}$ is the Fermi-Dirac distribution function, T is the temperature and $k_B$ = 8.6 × $10^{-5}$ eV/K is the Boltzmann's constant.

## Conflicts of interest

There are no conflicts to declare.

## Acknowledgements

This work was supported by the Engineering and Physical Sciences Research Council (APP17327).

## References

1 G. Sedghi, K. Sawada, L. J. Esdaile, M. Hoffmann, H. L. Anderson, D. Bethell, W. Haiss, S. J. Higgins and R. J. Nichols, *J. Am. Chem. Soc.*, 2008, **130**, 8582–8583.

2 H. Hao, H. Li, T. Jia, and X. Zheng. *Physical Chemistry Chemical Physics*, 2025, **27**, 331-339.

3 S. Sangtarash, A. Vezzoli, H. Sadeghi, N. Ferri, H. M. O'Brien, I. Grace, L. Bouffier, S. J. Higgins, R. J. Nichols and C. J. Lambert, *Nanoscale*, 2018, **10**, 3060–3067.

4 H. Sadeghi, S. Sangtarash and C. Lambert, *Nano Lett.*, 2017, **17**, 4611–4618.

5 I. V. Khariushin, P. Thielert, E. Zöllner, M. Mayländer, T. Quintes, S. Richert & A. Vargas Jentzsch, *Nature Chemistry*, 2025, 1-7.

6 H. Sadeghi, *Nanotechnology*, 2018, **29**, 373001.

7 D. J. Wold, R. Haag, M. A. Rampi and C. D. Frisbie, *J. Phys. Chem. B*, 2002, **106**, 2813–2816.

8 V. Kaliginedi, P. Moreno-Garcia, H. Valkenier, W. Hong, V. M. García-Suárez, P. Buiter, J. L. H. Otten, J. C. Hummelen, C. J. Lambert and T. Wandlowski, *J. Am. Chem. Soc.*, 2012, **134**, 5262–5275.

9 X. Zhao, C. Huang, M. Gulcur, A. S. Batsanov, M. Baghernejad, W. Hong, M. R. Bryce and T. Wandlowski, *Chem. Mater.*, 2013, **25**, 4340–4347.

10 H. C. Seong, B. Kim and C. D. Frisbie, *Science*, 2008, **320**, 1482–1486.

11 S. H. Choi, C. Risko, M. Carmen Ruiz Delgado, B. Kim, J. L. Brédas and C. Daniel Frisbie, *J. Am. Chem. Soc.*, 2010, **132**, 4358–4368.

12 C. Lungani Mthembu and R. C. Chiechi, *J. Mater. Chem. C,* 2025, **13**, 1272-1280.

13 N. Ramos-Berdullas and M. Mandado, *Chem. Eur. J.*, 2013, **19**, 3646–3654.

14 W. Xu, E. Leary, S. Sangtarash, M. Jirasek, M. T. González, K. E Christensen, L. Abellan Vicente, N. Agraït, S. J. Higgins, R. J. Nichols, C. J. Lambert, H. L. Anderson, *J. Am. Chem. Soc.* 2021, **143**, 20472–20481.

15 Y. Tsuji, R. Movassagh, S. Datta and R. Hoffmann, *ACS Nano*, 2015, **9**, 11109–11120.

16 T. Stuyver, S. Fias, F. De Proft and P. Geerlings, *Chem. Phys. Lett.*, 2015, **630**, 51–56.

17 Y. Zang, S. Ray, E. D. Fung, A. Borges, M. H. Garner, M. L. Steigerwald, G. C. Solomon, S. Patil and L. Venkataraman, *J. Am. Chem. Soc.*, 2018, **140**, 13167–13170.

18 N. Algethami, H. Sadeghi, S. Sangtarash and C. J. Lambert, *Nano Lett.*, 2018, **18**, 4482–4486.

19 E. Leary, B. Limburg, A. Alanazy, S. Sangtarash, I. Grace, K. Swada, L. J. Esdaile, M. Noori, M. T. González, G. Rubio-Bollinger, H. Sadeghi, A. Hodgson, N. Agraït, S. J. Higgins, C. J. Lambert, H. L. Anderson and R. J. Nichols, *J. Am. Chem. Soc.*, 2018, **140**, 12877–12883.

20 S. Sangtarash, C. Huang, H. Sadeghi, G. Sorohhov, J. Hauser, T. Wandlowski, W. Hong, S. Decurtins, S. X. Liu and C. J. Lambert, *J. Am. Chem. Soc.*, 2015, **137**, 11425–11431.

21 Y. Geng, S. Sangtarash, C. Huang, H. Sadeghi, Y. Fu, W. Hong, T. Wandlowski, S. Decurtins, C. J. Lambert and S. X. Liu, *J. Am. Chem. Soc.*, 2015, **137**, 4469–4476.

22 S. Reich, J. Maultzsch, C. Thomsen and P. Ordejón, *Phys. Rev. B - Condens. Matter Mater. Phys.*, 2002, **66**, 035412.

23 J. Ferrer, C. J. Lambert, V. M. García-Suárez, D. Z. Manrique, D. Visontai, L. Oroszlany, R. Rodríguez-Ferradás, I. Grace, S. W. D. Bailey, K. Gillemot, H. Sadeghi and L. A. Algharagholy, *New J. Phys.*, 2014, **16**, 093029.

24 J. M. Soler, E. Artacho, J. D. Gale, A. García, J. Junquera, P. Ordejón and D. Sánchez-Portal, *J. Phys. Condens. Matter*, 2002, **14**, 2745–2779.